\documentclass[lettersize,journal]{IEEEtran}

\ifCLASSOPTIONcompsoc
  \usepackage[nocompress]{cite}
\else
  \usepackage{cite}
\fi

\ifCLASSINFOpdf
\else
\fi

\usepackage[pdftex]{hyperref}

\usepackage[ruled]{algorithm2e}
\usepackage{my-tvcg}
\usepackage{epigraph}
\graphicspath{{./images/}}

\usepackage{color}
\definecolor{MainColor}{RGB}{128, 0, 128}

\hypersetup{
    unicode=true,          
    pdftoolbar=true,        
    pdfmenubar=true,        
    pdffitwindow=false,     
    pdfstartview={FitH},    
    pdftitle={Polynomial 2D Green Coordinates for High-order Cages},    
    pdfnewwindow=true,      
    colorlinks=true,       
    linkcolor=MainColor,          
    citecolor=MainColor,        
    filecolor=magenta,      
    urlcolor=MainColor,           
    breaklinks=MainColor
}

\usepackage[authoryear]{natbib}
\begin{document}

\title{Closed-form Cauchy Coordinates and Their Derivatives for 2D High-order Cages}

\author{Shibo~Liu,
        Ligang~Liu,
        Xiao-Ming~Fu
\IEEEcompsocitemizethanks
{
\IEEEcompsocthanksitem S. Liu, L. Liu, and X. Fu are with the School of Mathematical Sciences, University of Science and Technology of China.\protect\\
E-mail: aa1758926168@mail.ustc.edu.cn, lgliu@ustc.edu.cn, fuxm@ustc.edu.cn.
\IEEEcompsocthanksitem Corresponding author: Xiao-Ming~Fu.
}
}

\markboth{Journal of \LaTeX\ Class Files,~Vol.~14, No.~8, December~2019}%
{Shell \MakeLowercase{\textit{et al.}}: Bare Demo of IEEEtran.cls for Computer Society Journals}

\IEEEtitleabstractindextext{%
\begin{abstract}
We propose closed-form Cauchy coordinates and their derivatives for 2D closed high-order input cages composed of arbitrary-order polynomial curves. 
Our coordinates facilitate the transformation of input polynomial curves into output curves of any desired polynomial order. 
Central to our derivation is the creative use of the residue theorem with the logarithmic function to obtain the integral of a rational polynomial required for extending the classical 2D Cauchy coordinates to high-order input cages.
Our coordinates enable smooth cage-aware conformal deformations, and the derivatives allow for point-to-point deformation. 
Moreover, our derivation can be extended to the input cages with rational polynomial curves.
Through various 2D deformations, we demonstrate how users can intuitively manipulate \Bezier control points to achieve desired deformations easily.
\end{abstract}

}

\maketitle
\IEEEdisplaynontitleabstractindextext
\IEEEpeerreviewmaketitle

\section{Introduction} \label{sec:intro}
%
%

Cage coordinates offer powerful ways to define scalar or vector fields over 2D domains using sparsely specified boundary values (where the domain boundary, termed a \emph{cage}, encodes user-defined constraints). 
The coordinates have found broad utility in computer graphics applications. 
They enable intuitive control for tasks such as 2D shape deformation, image colorization, and artistic editing by propagating boundary constraints to the entire domain.

Existing cage coordinates can be classified into two primary categories based on their interpolation properties and control mechanisms. 
First, given user-specified values at each cage vertex, the value at any interior point is interpolated through a weighted combination of these vertex values, such as mean-value coordinates~\cite{Floater2003, Hormann2006}, positive mean-value coordinates \cite{Lipman2007}, and harmonic coordinates~\cite{Joshi2007}.
Second, other types of coordinates extend control beyond boundary values to include boundary derivatives (i.e., normals)
~\cite{Lipman2008, Weber2009, Hou2017, Hou2018, Ilbery2013, Weber2012}. 
Specifically, Green coordinates~\cite{Lipman2008} and complex-valued Cauchy coordinates~\cite{Weber2009} have been mathematically proven equivalent in 2D, with both formulations inherently producing conformal deformations.

 \begin{figure}
		\centering
		\begin{overpic}[width=0.99\linewidth]{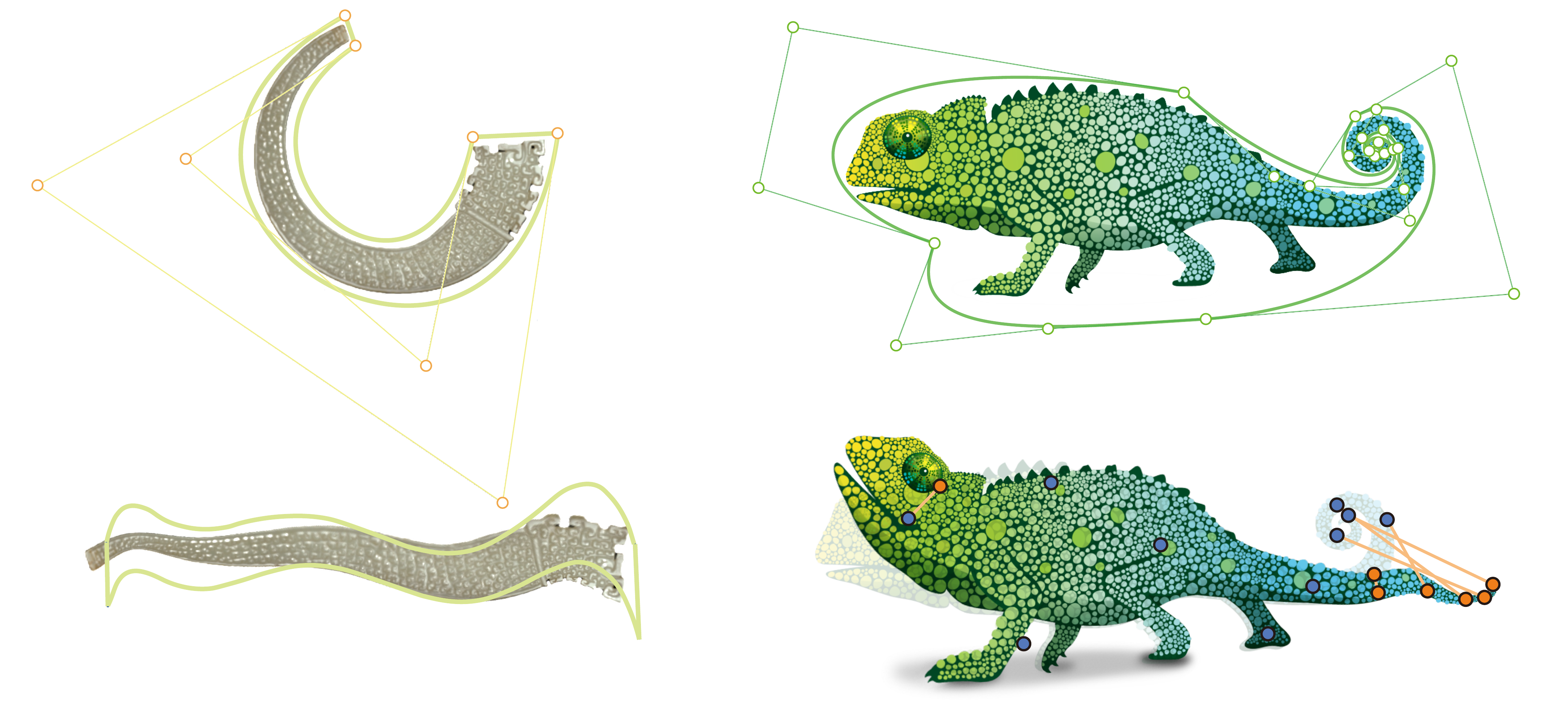}
			{
				  \put(10,20){\small Input }
                    \put(75,21){\small Input }
                    \put(15,2){\small Deformed }
                    \put(70,-2){\small Deformed }
                    %
		    }
		\end{overpic}
		 \vspace{-2mm}
		 \caption{
           Deformation via our coordinates and their derivatives. Upper: input images and cubic cages. 
           Bottom: cage-based deformation from cubic to 7th-order cages (left) and point-to-point deformation (right).
			 }
		 \label{fig:teaser}
		\end{figure}

\begin{figure}[!t]
  \centering
  \begin{overpic}[width=0.95\linewidth]{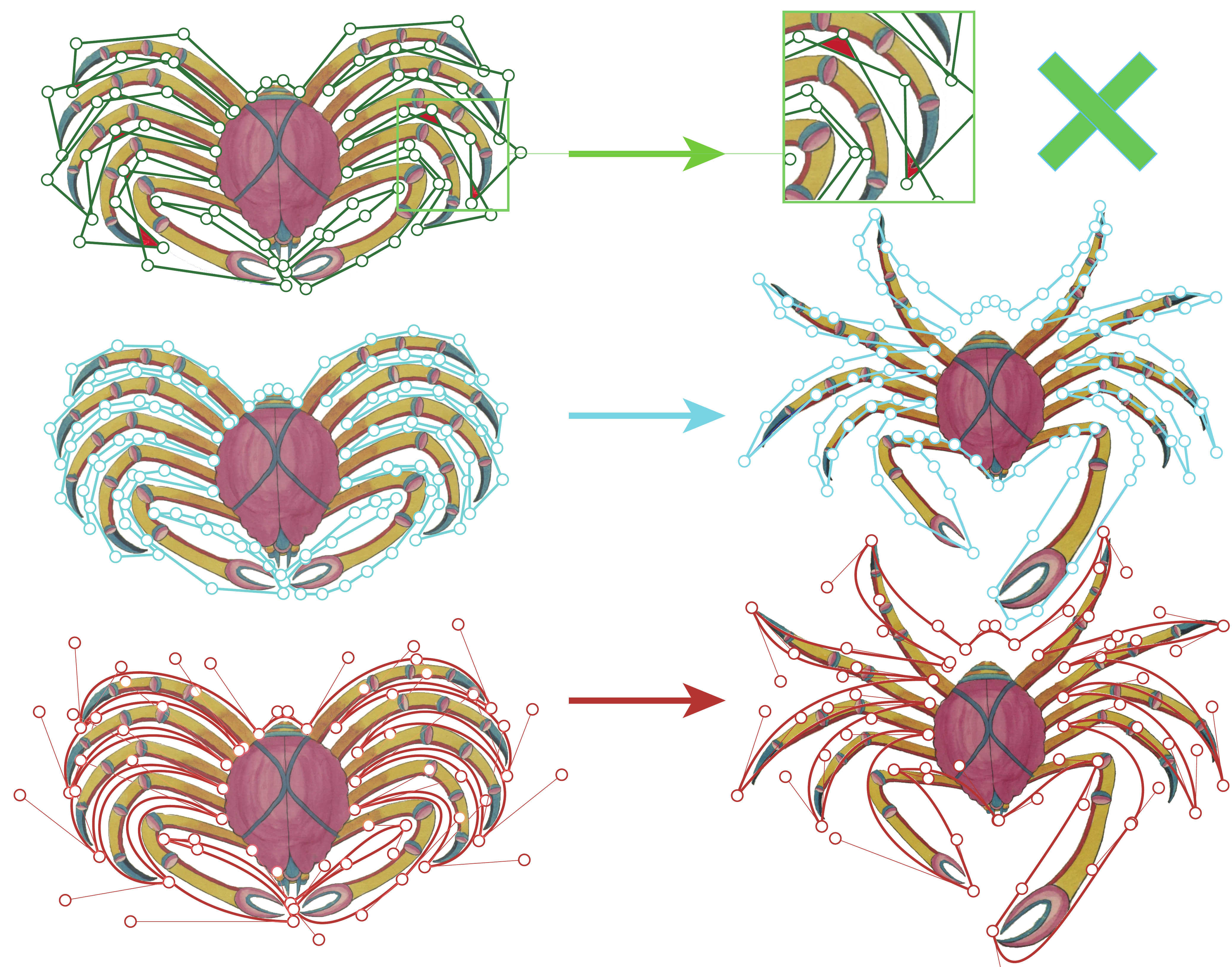}
    {
 \put(45,68){\small linear cage}
  \put(43,63){\small 72 Segments}
   \put(45,47){\small linear cage}
   \put(42,41.5){\small 120 Segments}
 \put(45,24){\small cubic cage}
   \put(44,18){\small 24 Segments}
    }
  \end{overpic}
  \vspace{-3mm}
  \caption{
Deformations using linear and high-order input cages. 
Upper: a cage with 72 linear segments exhibits self-intersections, preventing deformation.
Middle: deformation via a 120-segment polygonal cage under Cauchy coordinates.
Bottom: deformation with a 24-segment cubic cage using our coordinates.
High-order input cages lead to more intuitive deformations with many fewer parameters.
 }
  \label{fig:cmp-cages}
\end{figure}

\begin{figure}[!t]
  \centering
  \begin{overpic}[width=0.99\linewidth]{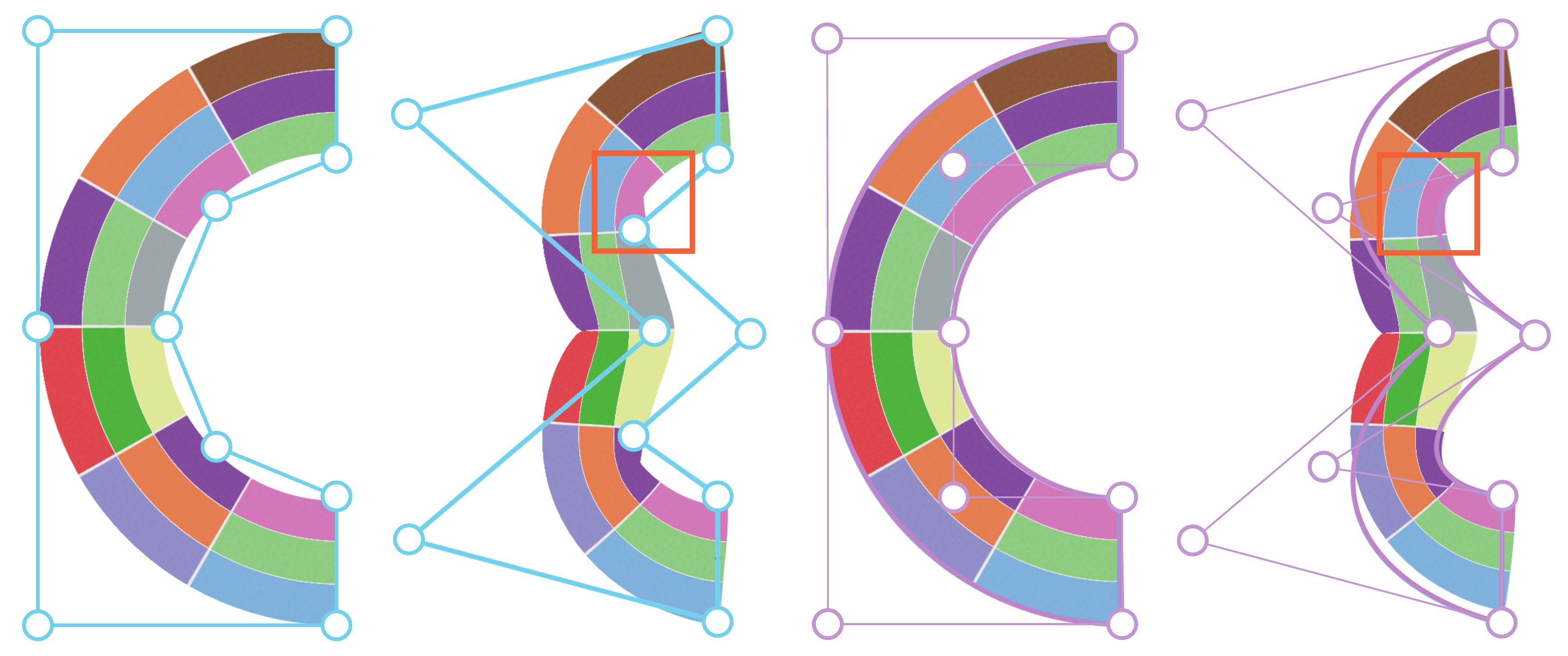}
    {
 \put(10,-1){\small \textbf{(a1)}}
 \put(35,-1){\small \textbf{(a2)}}
 \put(63,-1){\small \textbf{(b1)}}
 \put(90,-1){\small \textbf{(b2)}}
    }
  \end{overpic}
  \vspace{-2mm}
  \caption{
  Linear cage (a1) vs. quadratic rational cage (b1) with the same number of control points.
Deformation using our coordinates (b2) is smoother than the linear cage-based deformation via Cauchy coordinates (a2) (see the difference in the red boxes).}
  \label{fig:rational}
\end{figure}

Recent advances in cage coordinates, particularly cubic mean value coordinates~\cite{Li2013}, polynomial Green coordinates \cite{MichelThiery2023}, and polynomial Cauchy coordinates~\cite{Lin2024PolynomialCC}, have extended traditional approaches to enable cage-based deformation where initial straight cage segments are deformed into higher-order polynomial curves.
Addressing similar scenarios, \cite{Qin2024Coons} generalize arbitrary generalized barycentric coordinates through a novel parametric transfinite interpolation scheme.
However, these existing approaches are limited to input cages with straight edges.
Moving beyond conventional straight-edged cages, we investigate coordinates for cages composed of polynomial curves (e.g., represented via \Bezier curves), which we term \emph{high-order cages}.
%
%
High-order cages mainly offer three key benefits.  
First, compared to linear cages, they provide a higher-fidelity approximation to the original shape, leading to a more responsive and easier interaction for deformation while editing the cage (Fig.~\ref{fig:cmp-cages}). 
Second, the polynomial boundaries enable intuitive shape manipulation that parallels conventional free-form shape modeling in CAD systems and require fewer parameters for shape editing (Fig.~\ref{fig:cmp-cages}). 
Third, the deformations are smooth near boundaries (Figs.~\ref{fig:rational},~\ref{fig:cmp-smoothness}, and see Lemma~\ref{pro:lemma_holo}).
%
While \cite{Lin2024PolynomialCC} has extended Cauchy coordinates to high-order cages, it faces challenges. 
The method requires an intermediate straight-edged cage and relies on numerical computation of the inverse of the mapping from the intermediate to the input cages. 
These requirements cause: (1) the coordinates and their derivatives lack closed forms, failing to support point-to-point deformation, and (2) non-intuitive geometric constraints are imposed on high-order input cages, thus they cannot compute valid coordinates for arbitrary high-order input cages that enclose shapes.

In this paper, we propose the closed-form formulation of Cauchy coordinates and their derivatives for 2D high-order input cages.
The coordinates enable the deformation of input polynomial curves into output curves of arbitrary degree and produce conformal cage-based deformation. 
%
To address the primary challenge of computing the integral of rational polynomials to derive the closed forms, we creatively use the residue theorem instead of finding the antiderivative. Technically, we innovatively use the logarithmic function to constrain the integration domain and design corresponding integrands and integration paths to facilitate the application of the residue theorem.
Our residue theorem-based approach can compute the integral of an arbitrary rational polynomial, enabling our method to extend to the input cages with rational polynomial curves.

This analytical formulation enables rapid and accurate computation of Cauchy coordinates and their derivatives within high-order cages.
%
For 2D cage-based deformation, our coordinates preserve conformality while providing intuitive control, i.e., users can manipulate \Bezier control points to adjust polynomial curves for achieving desired deformations (Fig.~\ref{fig:teaser} - left).
Moreover, the derivatives make point-to-point (P2P) deformation possible (Fig.~\ref{fig:teaser} - right).


\section{Related work} \label{sec:related}
Numerous deformation techniques exist for images and 2D vector graphics, including free-form deformations~\cite{Sederberg1986}, moving-least-squares deformations~\cite{Schaefer2006}, and brush-based deformations~\cite{DeGoes2017}. 
Our review focuses specifically on 2D cage-based deformation approaches that are most relevant to our work. 

\paragraph{Cage coordinates}
Interpolating barycentric coordinates provide a mechanism for function interpolation within the cage domain.
Many coordinates have been proposed, such as mean-value coordinates (MVCs)~\cite{Floater2003,Hormann2006,Ju2005,Lipman2007}, harmonic coordinates~\cite{Joshi2007}, Poisson Coordinates~\cite{li2012poisson}, maximum entropy coordinates~\cite{Hormann2008Maximum}, maximum likelihood coordinates~\cite{Chang2023Maximum}, local barycentric coordinates~\cite{Zhang2014Local}, and variational barycentric
coordinates~\cite{Dodik2023Variational}. 
Except for the mean-value coordinates, other coordinates lack closed-form expressions, requiring numerical solutions. 
%

Non-interpolating coordinates achieve a trade-off between conformality and interpolation. 
\cite{Lipman2008} propose closed-form Green coordinates based on Green's third identity.
Green coordinates do not interpolate the deformation field on the cage boundary but preserve local geometric details.
In particular, they achieve conformal deformations in 2D.
The Cauchy coordinates~\cite{Weber2009} are shown to be equivalent to 2D Green coordinates. 
\cite{Weber2012} derive closed-form 2D biharmonic coordinates, providing control over boundary shapes and their derivatives. 
They significantly improve the geometric alignment between cages and shape boundaries.
%
\cite{thiery2024biharmonic} extend these coordinates to 3D.
The Somigliana coordinates~\cite{chen2023somigliana} are derived through a boundary integral formulation of linear elasticity, enabling volume control in 3D cage-based deformations.
We focus on deriving closed-form Cauchy coordinates and their derivatives for 2D high-order input cages.

\paragraph{Coordinates for high-order cages}
Cubic mean-value coordinates~\cite{Li2013} generalize traditional MVCs to allow deforming the cage edges into cubic curves. 
Subsequently, \cite{Qin2024Coons} further generalizes arbitrary generalized barycentric coordinates to enable the deformation of cage edges into curves of arbitrary degree through a novel parametric transfinite interpolation scheme.
Although these approaches enhance deformation flexibility while interpolating boundary positions, they sacrifice conformality.
Consequently, the resulting deformations exhibit merely continuity at cage vertices and fail to preserve local geometric features.
Polynomial 2D Green coordinates~\cite{MichelThiery2023} and Polynomial 2D Cauchy coordinates~\cite{Lin2024PolynomialCC} with closed-form solutions extend Green coordinates and Cauchy coordinates to represent a significant advancement by allowing the transformation of cage elements into polynomial curves of any degree. 
This extension maintains the conformal properties of traditional Green coordinates and Cauchy coordinates while offering enhanced control over the deformation and improved detail preservation. 

Moreover, while \cite{Li2013} demonstrate image editing using curved input cages, they require a preliminary straightening step to convert the curved cage into an intermediate straight cage and generate an intermediate image. This intermediate configuration then serves as the input to their deformation framework. 
Similarly, \cite{Lin2024PolynomialCC} also employ an intermediate linear cage
to compute the inverse of the mapping from the original cage,
which then enables the calculation of Cauchy coordinates for curved-edge cages.
In contrast, our coordinates are inherently for high-order input cages without any other processing. 


\paragraph{High-order meshes and cages}
High-order meshes have gained increasing attention in recent research, partly due to their excellent boundary-fitting capabilities~\cite{hu2019TriWild,mandad2020bezier,Mandad2021Guaranteed,Yang2022PreciseHM,Khanteimouri2022RationalBG,Jiang2021BijectiveAC,Khanteimouri20233Dbezier,liu2024curveshell}.
These meshes often consist of \Bezier triangles in 2D or \Bezier tetrahedrons in 3D~\cite{farin1986triangular}. 
%
%
Although the previous methods focus on generating cages with straight elements~\cite{Guo2024RobustCage,Sacht2015}, we can produce high-order cages by first evaluating their orders and then deforming to be curved while avoiding intersections, as done in~\cite{liu2024curveshell}.
It is necessary to develop coordinate systems for these high-order cages. 


%

%

\section{Method} \label{sec:method}
We review Cauchy coordinates in Sec.~\ref{sec:revisit}, then introduce Cauchy coordinates for high-order cages in Sec.~\ref{sec:High-order-Cauchy}, and finally derive the closed-form expressions of these coordinates and their derivatives in Sec.~\ref{sec::closed-form}.
More discussions are provided in Secs.~\ref{sec:connection} and~\ref{sec:details}.

\subsection{Revisiting Cauchy coordinates}\label{sec:revisit}
\paragraph{Cauchy transform}
Fundamental to complex analysis, Cauchy's integral formula provides that a holomorphic function on a planar domain \(\Omega\) with a simple closed boundary $\partial \Omega$ is completely determined by its values on $\partial \Omega$.
%
Substituting the boundary values of a continuous function \(f\) in the right-hand side of Cauchy's integral formula yields a holomorphic function \(u\): 
\begin{equation}\label{equ:Cauchy integral}
    u(z)=\frac{1}{2\pi i}\int_{\partial \Omega}\frac{f(\zeta)}{\zeta-z} d\zeta,\ \ z \in \Omega,
\end{equation}
where $z$ is a complex number and \(i=\sqrt{-1}\).
This transformation from \(f\) to \(u\) is called \emph{the Cauchy transform} \cite{Bell2015cauchytransform}.
The two Cauchy coordinates~\cite{Weber2009,Lin2024PolynomialCC} are from the Cauchy transform.

\begin{figure}[t]
  \centering
   \vspace{-2mm}
  \begin{overpic}[width=0.99\linewidth]{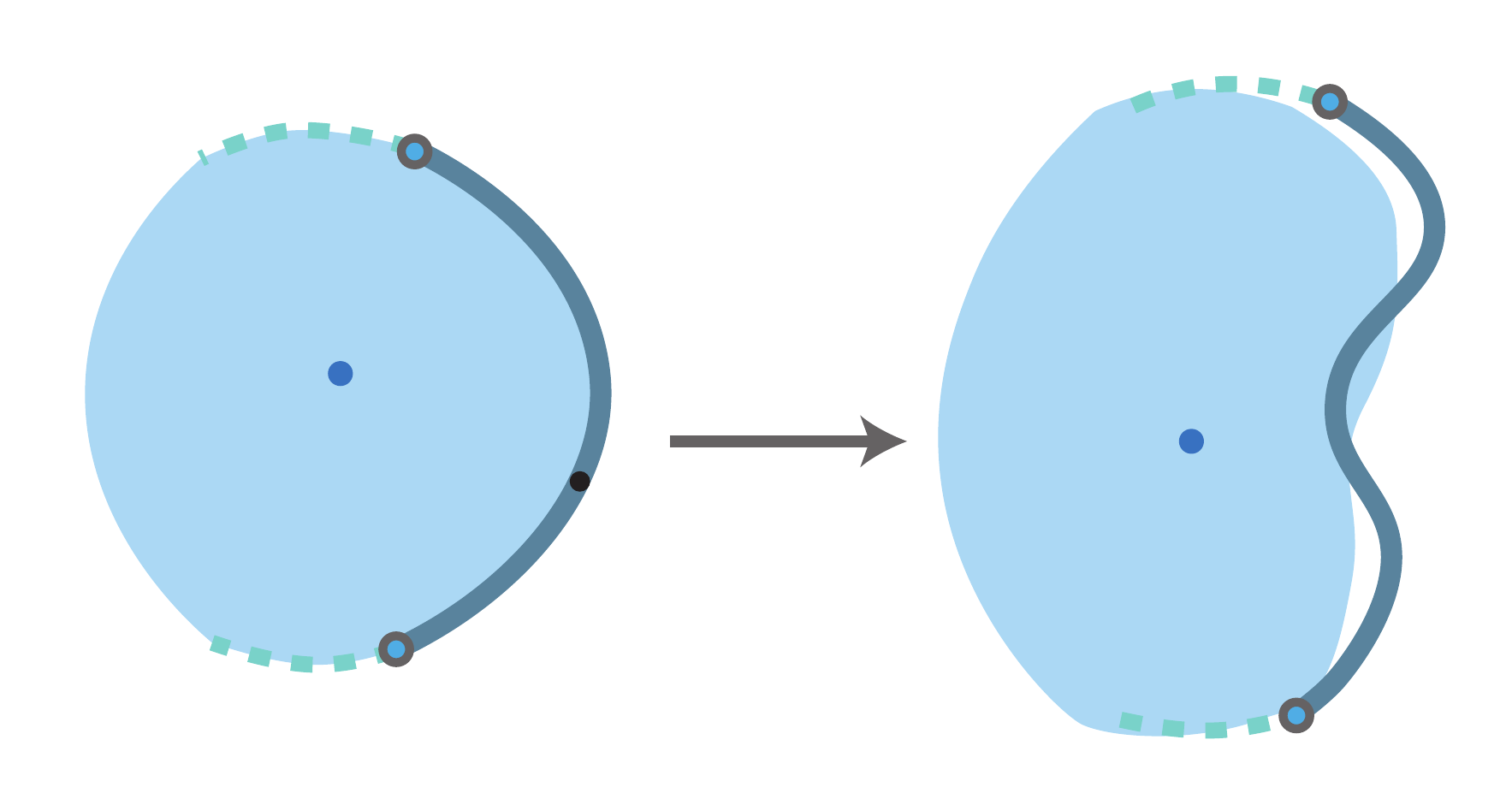}
    {
 \put(31,9){\small $\me_j$}
 \put(40,19){\small $\zeta$}
 \put(13,13.5){\small $\Omega$}
 \put(23,23.5){\small $z$}
 \put(50,24){\small $u$}
 \put(80,23.5){\small $u(z)$}
 \put(91,7){\small $f^{\text{new}}(\me_j)$}
    }
  \end{overpic}
  \vspace{-5mm}
  \caption{
  The deformation of a high-order cage. An original curve $\me_j$ is deformed into $f^{\text{new}(\me_j)}$ and any point $z$ inside the cage is deformed to $u(z)$.
  }
  \label{fig:notations}
\end{figure}

\paragraph{Cages}
Given the bounded domain $\Omega$, its boundary $\partial \Omega$, called a \emph{cage}, is non-intersecting and closed with a set of vertices $V = \{\mv_j\}^n_{j=1}$ connected by edges $E = \{\me_j\}^n_{j=1}$ (oriented counterclockwise) (Fig.~\ref{fig:notations}).
%
%
On the complex plane, each vertex $\mv_j$ is represented as a complex number \(z_j=\mv_{j,x}+\mv_{j,y} i\).
The Cauchy coordinates~\cite{Weber2009} and the polynomial Cauchy coordinates~\cite{Lin2024PolynomialCC} are defined on the rest cage with straight edges, i.e., \( z_{\me_j}(t) = (1-t) z_{j-1} + t z_j, t\in[0,1]\) for the edge $\me_j = \overline{\mv_{j-1}\mv_j}$; however, they differ in the deformed cage configurations: 
\begin{itemize}
    \item For Cauchy coordinates, the new $\me_j$ is linear, i.e., $f^\text{new}(z_{\me_j}(t)) = (1-t) p^\text{new}_{j-1}+t p^\text{new}_{j}$, where $p^\text{new}_{j}$ is the new position of $\mv_j$. 
    \item For the polynomial Cauchy coordinates, the new $\me_j$ becomes a polynomial curve of degree $n_j \geq 1$, i.e., $f^\text{new}(z_{\me_j}(t)) = \sum_{k=0}^{n_j} B^{n_j}_k(t) p^\text{new}_{\me_j,k}$, , where $B^{n_j}_k (t)$ is the Bernstein Polynomial and $p^\text{new}_{\me_j,k} \in \mathbb{C}$ is the control point.
\end{itemize}
Since the polynomial case includes the linear one, i.e., $n_j = 1$, we only discuss the polynomial Cauchy coordinates next.

\paragraph{Deriving Cauchy coordinates}
%
By substituting the new positions of the deformed cage into~\eqref{equ:Cauchy integral}, we obtain
\begin{equation}\label{equ:discretization}
\begin{split}
    u(z) &:=\frac{1}{2\pi i} \sum_{j=1}^n \int_{\me_j}\frac{f^\text{new}(\zeta)}{\zeta-z}d\zeta \\
     &=\frac{1}{2 \pi i}\sum_{j=1}^n \sum_{k=0}^{n_j} \left(\int_{\me_j}\frac{B^{n_j}_k (t)}{\zeta-z}d\zeta\right) p^\text{new}_{\me_j,k}, \ \ z \in \Omega.
\end{split}
\end{equation}

%
Users edit the control points $p^\text{new}_{\me_j,k}$ to deform the shape intuitively.
Moreover, the sum of all the integrals ($\frac{1}{2 \pi i}\left(\int_{\me_j}\frac{B^{n_j}_k (t)}{\zeta-z}d\zeta\right)$ in~\eqref{equ:discretization}) related to the control point $p^\text{new}_{\me_j,k}$ form its Cauchy coordinate~\cite{Lin2024PolynomialCC}. 
Since $B^{n_j}_k (t) = 
\begin{pmatrix}
n_j \\
k \\
\end{pmatrix}
t^k (1-t)^{n_j-k}$, 
we evaluate the integrals $\left(\int_{\me_j}\frac{t^k}{\zeta-z}d\zeta\right)$ with various $k$s to obtain the coordinates.


\subsection{2D Cauchy coordinates and their derivatives for high-order cages}\label{sec:High-order-Cauchy}

\paragraph{Enhanced deformation interactivity}
According to Cauchy's integral formula, we have
\begin{equation}
    u(z)-z=\frac{1}{2\pi i}\int_{\partial \Omega}\frac{f(\zeta)-\zeta}{\zeta-z} d\zeta,\ \ z \in \Omega,
\end{equation}
where \(f(\zeta)-\zeta\) represents the displacement of the point $\zeta$ on the cage (which is typically not holomorphic) and \(u(z)-z\) is the deformation displacement of \(z\). Since \(u(z)-z\) is a holomorphic function, it follows from the maximum modulus principle that the maximum deformation displacement occurs on \(\partial \Omega\). 
Consequently, when applying a displacement field to the cage, shapes closer to the cage experience greater deformation displacement, resulting in more responsive deformation near cage boundaries. 
For effective deformation, the cage should closely conform to the shape rather than enclosing it within a loose, distant boundary.
High-order cages offer superior approximation capabilities to linear cages, motivating our generalization of Cauchy coordinates to high-order cages.

\paragraph{Holomorphicity on the boundary}
Both Cauchy coordinates and polynomial Cauchy coordinates yield a function $u$ that is holomorphic inside \(\Omega\). 
Since holomorphic functions with non-zero derivatives are conformal mappings, they produce conformal deformation. 
However, $u$ is not holomorphic on \(\partial \Omega\), and this non-holomorphicity leads to non-smooth results near \(\partial \Omega\) using Cauchy coordinates (see Fig.~4 in \cite{Lin2024PolynomialCC}). 
If the input 2D shape consists only of straight edges, it can be closely approximated by an initial linear cage. In that case, polynomial Cauchy coordinates address this problem since they deform straight edges into \Bezier curves to obtain smooth results according to the following theorem.
 \begin{theorem}[Theorem 3.3 in \cite{Bell2015cauchytransform}]\label{pro:holomorphic}
Suppose that \(\partial \Omega\) is smooth and $u$ is a holomorphic function on $\Omega$ that extends to be a continuous function on $\Omega\cup \partial \Omega$. If the boundary values of $u$ are smooth on an open arc $\Gamma$ in $\partial\Omega$, then all the derivatives of $u$ extend continuously from $\Omega$ to $\Omega \cup \Gamma$.
\end{theorem}

\begin{figure}[t]
  \centering
  \begin{overpic}[width=0.99\linewidth]{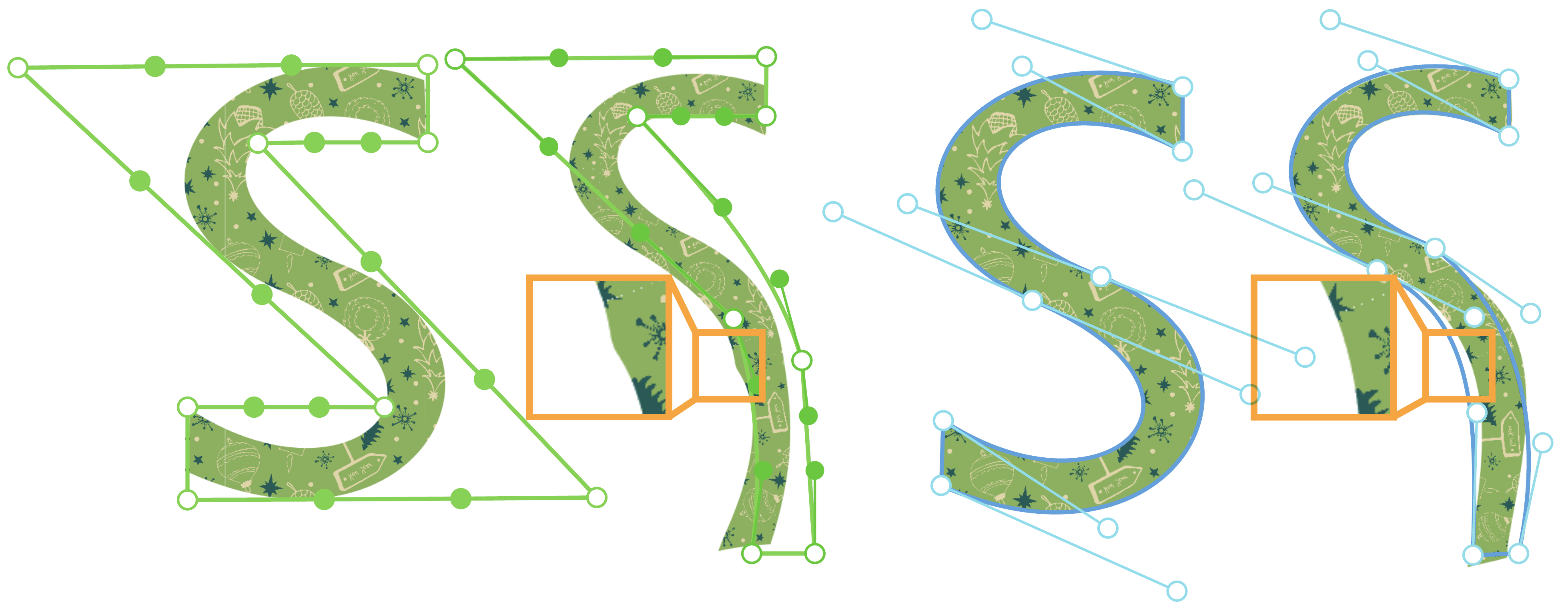} %
    {
 \put(16,-2){\small \textbf{(a1)}}
 \put(40,-2){\small \textbf{(a2)}}
 \put(67,-2){\small \textbf{(b1)}}
 \put(89,-2){\small \textbf{(b2)}}
    }
  \end{overpic}
  \vspace{-1mm}
  \caption{
  Polynomial Cauchy coordinates mapping a linear cage (a1, with non-endpoint control points on edges shown as solid dots) to a cubic cage lead to non-smoothness near the boundaries (a2).
  Our cubic input cage tightly encloses the shape (b1), generating smooth deformation (b2).
  }
  \label{fig:cmp-smoothness}
\end{figure}

Nevertheless, the problem of non-holomorphicity on the boundary still exists when a linear input cage is used for a shape with curved boundaries (Fig.~\ref{fig:cmp-smoothness}).
Fortunately, Theorem~\ref{pro:holomorphic} also indicates that the holomorphicity of $u$ on a specific boundary edge is guaranteed if the edges before and after deformation are smooth. This motivates us to generalize Cauchy coordinates to high-order input cages, which can tightly enclose the input shapes using smooth edges. 
Assuming that \(\partial \Omega\) is a piecewise smooth boundary, we can readily deduce the following lemma from Theorem~\ref{pro:holomorphic}: 
 \begin{lemma}\label{pro:lemma_holo}
 The holomorphic function \(u\) on \(\Omega\) is holomorphic on \(\partial \Omega\) if and only if \(f\) maps smooth segments of the boundary to smooth segments, and \(f\) acts as a similarity transformation at the non-smooth vertices.
 \end{lemma}


\paragraph{Derivation}
We now extend the Cauchy coordinates to a rest high-order cage $\partial \Omega$ formed by a set of polynomial curves.
The input edge $\me_j$ is represented a \Bezier curve $z_{\me_j}(t)$ of degree $m_j \geq 1$: $z_{\me_j}(t) = \sum^{m_j}_{k=0} B_k^{m_j}(t) p^\text{old}_{\me_j,k} , \forall t \in [0, 1]$, where $p^\text{old}_{\me_j,k} \in \mathbb{C}$ is the control point. 
Without loss of generality, we still assume that the deformed curve of $\me_j$ is $f^\text{new}(z_{\me_j}(t)) = \sum_{k=0}^{n_j} B^{n_j}_k(t) p^\text{new}_{\me_j,k}$.
Then, we have
\begin{equation}\label{equ:discrete-curved}
\begin{split}
    u(z) &=\frac{1}{2\pi i} \sum_{j=1}^n \int_{\me_j}\frac{f^\text{new}(\zeta)}{\zeta-z}d\zeta \\
    &= \frac{1}{2\pi i} \sum_{j=1}^n \int^1_{0} \frac{\sum_{k=0}^{n_j} p^\text{new}_{\me_j,k} B^{n_j}_k(t) z'_{\me_j}(t)}{z_{\me_j}(t) - z} dt \\
     &= \frac{1}{2\pi i} \sum_{j=1}^n \sum_{k=0}^{n_j} \left(\int^1_{0} \frac{ B^{n_j}_k(t) z'_{\me_j}(t)}{z_{\me_j}(t) - z} dt \right) p^\text{new}_{\me_j,k}, \ \ z \in \Omega.
\end{split}
\end{equation}
We denote the integral of~\eqref{equ:discrete-curved} in the bracket as $\alpha_{\me_j,k}(z)$.
All the integrals related to the control point $p^\text{new}_{\me_j,k}$ form its Cauchy coordinate.

%
%
The $l$th-order derivative of $\alpha_{\me_j,k}(z)$ is
\begin{equation}\label{equ:derivative-curved}
     \frac{d^l \alpha_{\me_j,k}(z))}{ dz^l } = l! \int^1_{0} \frac{ B^{n_j}_k(t) z'_{\me_j}(t)}{(z_{\me_j}(t) - z)^{l+1}} dt.
\end{equation}
This derivative enables easy computation of coordinate derivatives.
%

\subsection{Closed-form solutions}\label{sec::closed-form}
The integrand in the integral of~\eqref{equ:discrete-curved} for coordinate or~\eqref{equ:derivative-curved} for coordinate derivative is a rational polynomial. 
Then, we present a method to compute an analytic solution of this kind of integral.
This is our core contribution.

\paragraph{Representation}
The denominator $h(t), t\in[0,1]$ of the rational polynomial is a polynomial of degree $l$.
The numerator is represented as a polynomial of degree $r$: $g(t) = \sum^r_{k=0} a_k t^k$, where $a_k$ is the coefficient.
Since $\int^1_{0} g(t) / h(t) dt = \sum^r_{k=0} a_k \int^1_{0} t^k / h(t) dt$, we compute $\int^1_{0} t^k / h(t) dt, \forall k \geq 0$ to achieve the integral. 


\paragraph{Theorem}
The following theorem provides the result.
\begin{theorem}\label{pro:general}
The polynomial $h(t)$ has $m\leq l$ different roots $\{\zeta_j\}^m_{j=1}$ ($\zeta_j$ is a complex number).
Then, 
$$\int_0^1 \frac{t^k}{ h(t)}dt=\sum_{j=1}^{m}\res(S_k(\zeta),\zeta_j), \zeta \in \mathbb{C}, $$
where $S_k(\zeta)=s_k(\zeta)/h(\zeta)$ and
$$s_k(\zeta)=\zeta^k(\log(1-\frac{1}{\zeta})+\sum_{j=1}^k\frac{1}{j \zeta^j}).$$
\end{theorem}


\begin{figure}[!t]
  \centering
  \begin{overpic}[width=0.5\linewidth]{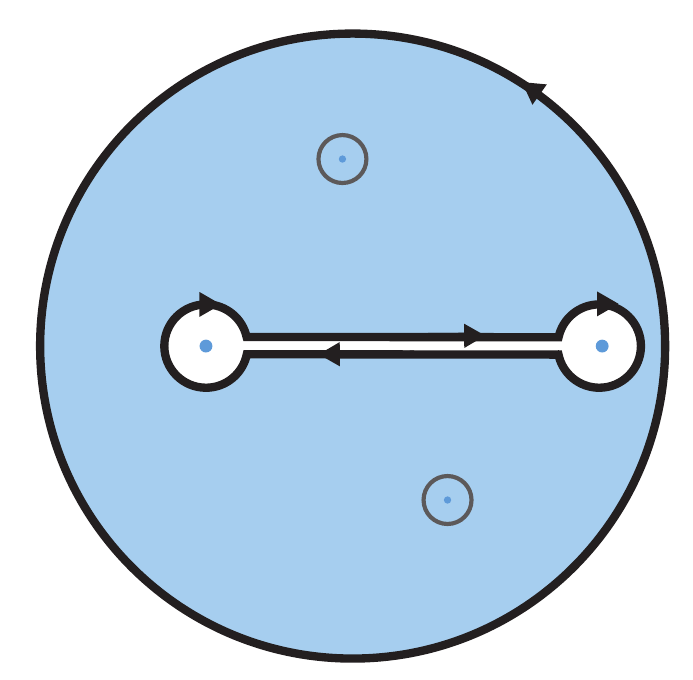}
    {
 \put(63,19){\small $\zeta_i$}
 \put(52,70){\small $\zeta_j$}
 \put(65,54){\small $l_1$}
 \put(50,40){\small $l_2$}
 \put(27,36){\small $0$}
 \put(85,36){\small $1$}
 \put(85,58){\small $\gamma_2$}
 \put(27,58){\small $\gamma_1$}
 \put(84,80){\small $\Gamma$}
    }
  \end{overpic}
  \vspace{-3mm}
  \caption{
The complex path taken for the integration.
  }
  \label{fig:inte_path}
\end{figure}

\begin{proof}
We consider the integral of function \(S_k(\zeta)\) along the integration path (see Fig.~\ref{fig:inte_path}) composed of \(\gamma_1,l_1,\gamma_2,l_2,\) and \(\Gamma\), where \(\gamma_1,\gamma_2\) are circles of radius \(\epsilon\) centered at 0 and 1 respectively, \(l_1\) and \(l_2\) are constitute the upper and lower banks of the line segment \([0,1]\), and \(\Gamma\) is a large circle of radius \(R\) (\(R \to \infty\) in this proof). 

\(S_k(\zeta)\) is holomorphic on this integration path. In fact, \(\log\left(1 - \frac{1}{\zeta}\right) = \log(1 - \zeta) - \log(-\zeta)\). When \(\zeta\) makes one full turn along any simple closed curve outside the line segment \([0,1]\), $\log\left(1 - \frac{1}{\zeta}\right)$ remains a single-valued holomorphic branch.
By applying the \emph{residue theorem} to $S_k$ in the region enclosed by the path, we obtain:
    \begin{equation}\label{equ:residue_theorem}
        \begin{aligned}
 \int_{\gamma_1+l_1+\gamma_2+l_2+\Gamma} S_k(\zeta) \, d\zeta
            = 2\pi i \sum_{j=1}^{m} \res(S_k(\zeta), \zeta_j).
        \end{aligned}
    \end{equation}
When $\zeta\in l_1$, $\log(1-\zeta)-\log(-\zeta)=\log(1-t)+2 \pi i-\log(-t)$,
and when $\zeta\in l_2$, $\log(1-\zeta)-\log(-\zeta)=\log(1-t)-\log(-t)$, where $t$ is the real part of $\zeta$.
It follows that:
\begin{equation}\label{equ:l1l2}
\begin{aligned}
    &\int\limits_{l_1+l_2} S_k(\zeta) \, d\zeta=\int_{0}^1 \left(S_k(t) +\frac{2\pi i t^k}{h(t)}\right)\, dt+\int_1^0 S_k(t) \, dt\\
    &=2\pi i \int_{0}^{1} \frac{ t^k}{h(t)} \, dt.
    \end{aligned}
\end{equation}
 When \( \zeta \in \Gamma \), according to the Taylor expansion, \( s_k \) becomes:
    \begin{equation}
        s_k(\zeta)=\zeta^k(\sum_{j=1}^{\infty}\frac{-1}{j \zeta^j}+\sum_{j=1}^k\frac{1}{j \zeta^j})=\sum_{j=1}^\infty\frac{-1}{(k+j) \zeta^{j}}.
    \end{equation}
    Since the degree of $h(\zeta)$ is at least 1, we have $\lim_{R\to \infty}\zeta S_k(\zeta)=0$, which shows that 
    \begin{equation}\label{equ:Gamma}
        \lim\limits_{R\to \infty}\int_{\Gamma} S_k(\zeta) \, d\zeta=0.
    \end{equation}   
    When $\zeta \in \gamma_1$, we have:
    \begin{equation}
    \begin{aligned}
       \lim\limits_{\epsilon\to 0} &\int_{\gamma_1} S_k(\zeta) \, d\zeta=\lim\limits_{\epsilon\to 0}\int_{\gamma_1} \frac{\zeta^k \log(-\zeta)}{h(\zeta)}\, d\zeta.\\
       &\leq \frac{\epsilon^k (-\log(\epsilon)+2 \pi)}{h(0)}2 \pi \epsilon \to 0.
       \end{aligned}
    \end{equation}
    It is similar for $\zeta \in \gamma_2$. So we have:
    \begin{equation}\label{equ:gamma12}
       \lim\limits_{\epsilon\to 0} \int_{\gamma_1+\gamma_2} S_k(\zeta) \, d\zeta=0.
    \end{equation}
    
    After substituting~\eqref{equ:l1l2},~\eqref{equ:Gamma}, and~\eqref{equ:gamma12} into~\eqref{equ:residue_theorem}, we obtain:
    \begin{equation}
        \int_0^1 \frac{t^k}{ h(t)}dx=\sum_{j=1}^{m}\res(S_k(\zeta),\zeta_j).
    \end{equation}
\end{proof}

\paragraph{Remarks} 
In our proof, the term \(\log(1-\zeta) - \log(0-\zeta)\) restricts the integral to the interval \([0,1]\), while the additional term \(\sum \frac{1}{j \zeta^j}\) ensure that the residue of \( S_k \) at infinity is zero. 
The theorem can also be extended to compute integrals over \([a,b]\), as long as the logarithmic term in $s_k(\zeta)$ is modified to \(\log(b-\zeta) - \log(a-\zeta)\), and the additional term remains the negative of the first $k$ terms of the Taylor expansion of the logarithmic term. 


\paragraph{General roots}
In the most general case, \( \zeta_j \) is simple pole of \( S_k(\zeta) \). Consequently, if we denote $h^{j,1}(\zeta)=h(\zeta)/(\zeta-\zeta_j)$, we have:
\begin{equation}\label{equ:gene_expression}
\begin{aligned}
    &\res(S_k(\zeta),\zeta_j)=\lim_{\zeta\to \zeta_j}(\zeta-\zeta_j)S_k(\zeta)=\frac{s_k(\zeta_j)}{h^{j,1}(\zeta_j)}.
    \end{aligned}
\end{equation}

\paragraph{Repeated roots}
Multiple poles will appear at a few points when calculating \(\alpha_{\me_j,k}(z)\). For instance, if the input edge $\me_j$ is quadratic, the presence of multiple poles signifies that \(z\) coincides with the focus of \(\me_j\).
%
In this case, we denote \( \zeta_j \) as an \( l_j \)-th order pole and \(h^{j,l_j}(\zeta)=h(\zeta)/(\zeta-\zeta_j)^{l_j}\). Then, we have:
\begin{equation}\label{equ:repeated_expression}
\begin{aligned}
&\res(S_k(\zeta),\zeta_j) 
    =\lim_{\zeta\to \zeta_j}\frac{1}{(l_j-1)!}\frac{d^{l_j-1}}{d\zeta^{l_j-1}}\{\frac{s_k(\zeta)}{h^{j,l_j}(\zeta)}\}.
    \end{aligned}
\end{equation}
Applying standard differentiation rules, its computation is straightforward for a given~$l_j$.
In practice we haven't encountered cases where $l_j > 1$ when calculating \(\alpha_{\me_j,k}(z)\).



\subsection{Extensions}\label{sec:connection}
\paragraph{Polynomial Green coordinates}
\citet{MichelThiery2023} propose the closed-form polynomial Green coordinates to deform straight-edge input cages into curved ones; however, they can not handle high-order input cages.
We extend their work to derive Green coordinates for high-order cages using the proposed theorem.
According to Equation (12) in~\cite{MichelThiery2023}, the computation of Green coordinates for high-order cages reduces to evaluating an integral of the following form:
\[\int_{0}^1 \frac{t^k}{2\pi \|z-z_{\me_j}(t)\|^2}dt, z \in \Omega.\]
Theorem~\ref{pro:general} can be applied to this integral. 

\paragraph{Rational polynomial cages}
We can extend our Cauchy coordinates to a rest high-order cage $\partial \Omega$ formed by rational polynomial curves (Fig.~\ref{fig:rational}). Let $z_{\me_j}(t) = \frac{z_{w,\me_j}(t)}{w(t)}, \forall t \in [0, 1]$ , where \(z_{w,\me_j}(t)=\sum_{k=0}^{m_j}w_{\me_j,k}p_{\me_j,k}^{\text{old}}B_k^{n_j}(t)\) and \(w(t)=\sum_{k=0}^{m_j}w_{\me_j,k}B_k^{n_j}(t)\). We then express $f^\text{new}(z_{\me_j}(t))$ as a rational curve that shares the same weights but has new control points \(p^\text{new}_{\me_j,k}\), leading to:
\begin{equation}
    \alpha_{\me_j,k}(z)=\int_0^1 \frac{z_{w,\me_j}'(t)w(t)-w'(t)z_{w,\me_j}(t)}{(z_{w,\me_j}(t)-zw(t))w(t)^2}dt.
\end{equation}
Theorem~\ref{pro:general} also applies to this integral. We must account for scenarios where the roots of \(w(t)\) and \(z_{\me_j}^{w}(t)-zw(t)\) are identical. In these instances, we will consolidate the roots subject to an error tolerance \(\epsilon=10^{-10}\) and then re-evaluate their multiplicities.


\subsection{Implementation details}\label{sec:details}
Alg.~\ref{alg:encoding} gives the pseudocode for computing our Cauchy coordinates. 

\paragraph{Calculating \(\{\zeta_j\}\)}
To apply Theorem~\ref{pro:general} for obtaining Cauchy coordinates within high-order cages, we need to solve roots $\{\zeta_j\}$ of $h(t) = 0$. 
Closed-form root-finding formulas exist for the most widely used quadratic and cubic curves.
Numerical methods are necessary to find the roots of a polynomial of degree above four.

\paragraph{Recursion}\label{sec:encoding}
We calculate \(\res(S_{k}(\zeta),\zeta_j)\) iteratively. For simplicity, we will only present the case where \(\zeta_j\) is a simple pole; the case for multiple poles is similar. Since
    $s_{k+1}(\zeta)=\zeta s_k(\zeta)+\frac{1}{k+1},\ k\geq 0,$
we use the following recursive form instead in practice: 
\begin{equation}\label{equ:recur}
\begin{aligned}
&\res(S_{0}(\zeta),\zeta_j)=\frac{s_0(\zeta_j)}{h^{j,1}(\zeta_j)},\\
    &\res(S_{k+1}(\zeta),\zeta_j)=\zeta_j \res(S_k(\zeta),\zeta_j)+\frac{1}{k+1}\frac{1}{h^{j,1}(\zeta_j)}.
    \end{aligned}
\end{equation}

\begin{algorithm}[t]
  \caption{Cauchy coordinates of a rest position $z$}\label{alg:encoding}
  \SetKwInput{KwInput}{Input}
  \SetKwInput{KwOutput}{Output}
  \SetKwFunction{FMain}{Main}
  \SetKwFunction{FD}{D}
  
  \KwInput{A point $z$ inside the rest cage and an edge $\me_j$ of the rest cage represented by a polynomial $z_{\me_j}(t)$ of degree $m_j$.}
  \KwOutput{The coordinate $\alpha_{\me_j,r}(z)$ of $z$ with respect to control points \(p_r^{\text{new}}\) of deformed edge.}
  Calculate roots $\{\zeta_j\}_{j=1}^l$ of $z_{\me_j}(t)-z=0$\;
  Convert \(B_k^{n_j}(t)z_{\me_j}'(t)\) to the form \(\sum_{k=0}^{m_j+n_j-1} c_k t^k\)\;
  $k \gets 0; \alpha_{\me_j,r}(z) \gets 0$\;
  \For{$k \leq m_j+n_j-1$}{
  $temp \gets 0$\;
      \For{each \(\zeta_j\) in \(\{\zeta_j\}\)}{
      \If{\(\zeta_j\) is a simple pole}{Compute \(\res(S_{k}(\zeta),\zeta_j)\) using \eqref{equ:recur}.}
      \Else{Compute \(\res(S_{k}(\zeta),\zeta_j)\) using \eqref{equ:repeated_expression}.}
      $temp \gets temp+\res(S_k(\zeta),\zeta_j))$;
      }
      \(\alpha_{\me_j,r}(z) \gets \alpha_{\me_j,r}(z) + c_k temp; \, k \gets k+1 \)\;
  }
\end{algorithm}

\paragraph{Numerical Stability} 
When using~\eqref{equ:recur} for computation, we only need to compute the terms $1/h^{j,1}(\zeta_j)$ and $s_0(\zeta_j)$. $1/h^{j,1}(\zeta_j)$ is well defined when \(\zeta_j\) is a simple root.
The function \( s_0(\zeta) \) is given by:
\[{\small
\log(1-\frac{1}{\zeta}) = \frac{1}{2} \log\left(1+\frac{ 1 - 2 \operatorname{Re}(\zeta)}{\|\zeta\|^2}\right) + i \text{atan2}\left(\operatorname{Im}(\zeta), \|\zeta\|^2 - \operatorname{Re}(\zeta)\right).}
\]
It is well-defined for \( z \notin \partial \Omega \) since (1) the logarithmic function \( \log(\cdot) \) is ill-defined only for \( \|\zeta\|^2 = 0 \) or \( 1 - 2 \text{Re}(\zeta) \leq -\|\zeta\|^2 \), which is equivalent to \( \zeta = 0 \) or \( \zeta = 1 \) (i.e., for \( z \) on the extremities of \( \me_j \)) and (2) the arctangent function is well-defined for \( \zeta \neq 0 \) or \( \zeta \neq 1 \) as well.
We check the numerical errors of our coordinates. In all the examples, the maximum error never exceeded $10^{-11}$.

\section{Experiments and evaluations}\label{sec:results}
Our method is implemented in C++. 
All experiments were conducted on a desktop PC with a 4.00 GHz Intel Core i7-4790K processor and 16 GB of RAM.
Except for Fig.~\ref{fig:rational}, the edges of the high-order input cages in other figures are polynomials.

\paragraph{Cage-based deformation}
We show various examples of image deformation in Fig.~\ref{fig:more-example}. 
Our coordinates provide smooth and conformal deformations for organic and mechanical images. 
Our method can also go higher and lower in the polynomial degrees of the curved segments (Fig.~\ref{fig:diff-order}).
In the case where the output degree is less than the input, reducing the degree of the curves initially alters the shape of the cage without affecting the artist's subsequent edits. 
Moreover, Lemma~\ref{pro:lemma_holo} guarantees that the resulting deformation maintains smoothness near the edges.

\paragraph{Point-to-point (P2P) deformation}
To enable handle-driven deformation, we utilize coordinate derivatives to define an objective function optimized to obtain the deformation, similar to \cite{Weber2009} and \cite{Lin2024PolynomialCC}.
The objective function includes a sum of the squared distances between handles and a smoothness term composed of the $\ell_2$-norm of the second-order derivatives of the coordinates on the boundary. 
This optimization is a least-squares problem. A linear solve is applied to obtain the minimizer. Moreover, the coefficient matrix of the linear system is fixed during moving handles; thus, we pre-factorize it once during the preprocessing step, allowing editing handles interactively (Figs.~\ref{fig:teaser},~\ref{fig:more-example}, and \ref{fig:diff-order}).

\paragraph{Time and memory}
All cage-based and P2P deformations run in real-time in all demonstrated examples.
We analyze only pre-computation time.
As detailed in Sec.~\ref{sec:details}, once the roots \(\{\zeta_j\}\) are determined, evaluating the analytical expression for the integral involves only a few complex multiplications and logarithmic operations. 
In practice, computing \(\{\zeta_j\}\) with closed-form expressions takes much pre-computation time (approximately 24\% for quadratic curves and 33\% for cubic curves).
%
Thus, once \(\{\zeta_j\}\) is computed, it is convenient to first compute the integrals of rational polynomials up to a higher specified degree. 
%
This method lets artists freely adjust curve editing degrees up to the specified maximum without recomputing coordinates.
In our tests, preprocessing uses $\leq 90$ MB memory, a benchmark achieved by the Crab model in Fig.~\ref{fig:cmp-cages}, whose cage contains 24 cubic curves and encloses 27000 points internally.

\paragraph{Comparisons to~\cite{Weber2009}}
Cauchy coordinates~\cite{Weber2009} transform linear cages to linear cages.
Linear cages typically require excessive degrees of freedom to represent complex curved shapes, significantly compromising interactive manipulation (Figs.~\ref{fig:cmp-cages} and~\ref{fig:cmp-linearcages}). 
In contrast, the curved cage produces more natural deformations with fewer controls, enabling simpler editing with fewer adjustments.
For the examples in Fig.~\ref{fig:cmp-cages}, linear-cage approaches require significantly more complex operations and greater time to match high-order-cage results.
%
Moreover, gaps exist between linear cages and complex shapes, leading to non-smooth deformations near cages (see cage-based deformation in Fig.~\ref{fig:rational} and P2P deformation in Fig.~\ref{fig:comparep2p}). In contrast, the high-order cages maintain high geometric conformity, leading to smooth deformation. 

\begin{figure}[t]
  \centering
  \begin{overpic}[width=0.95\linewidth]{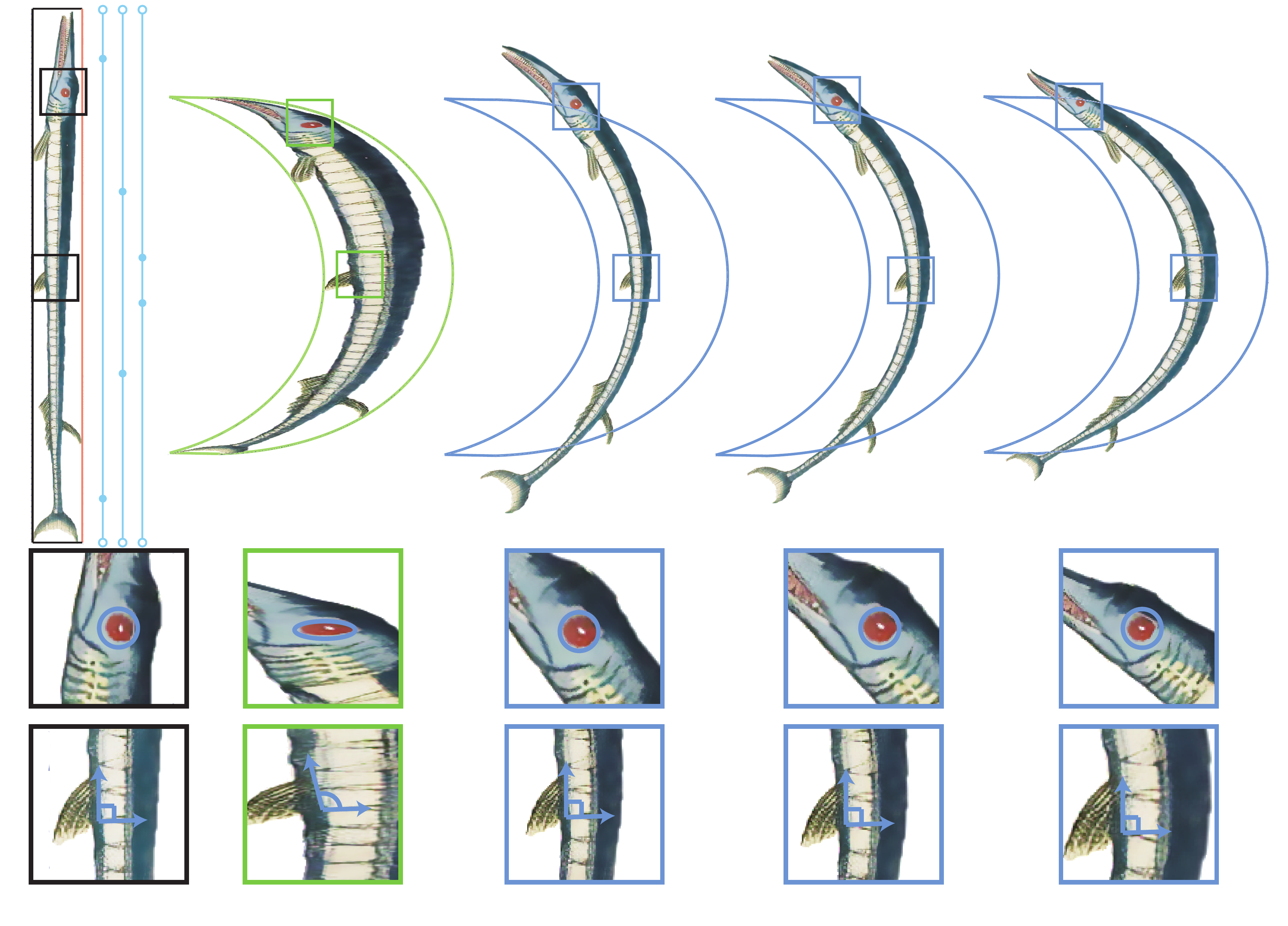} %
    {
    \put(7,-0.5){\small \textbf{(a)}}
 \put(23,-0.5){\small \textbf{(b)}}
 \put(43,-0.5){\small \textbf{(c1)}}
 \put(63,-0.5){\small \textbf{(c2)}}
 \put(85,-0.5){\small \textbf{(c3)}}
    }
  \end{overpic}
  \vspace{-2mm}
  \caption{
  (a) The input linear cage.
  (b) Cubic MVC.
  (c1, c2, c3) Our deformation results with three different initial parameterizations for the input edge (a).
  }
  \label{fig:cmp-mvc}
\end{figure}

\begin{figure}[t]
  \centering
  \begin{overpic}[width=0.95\linewidth]{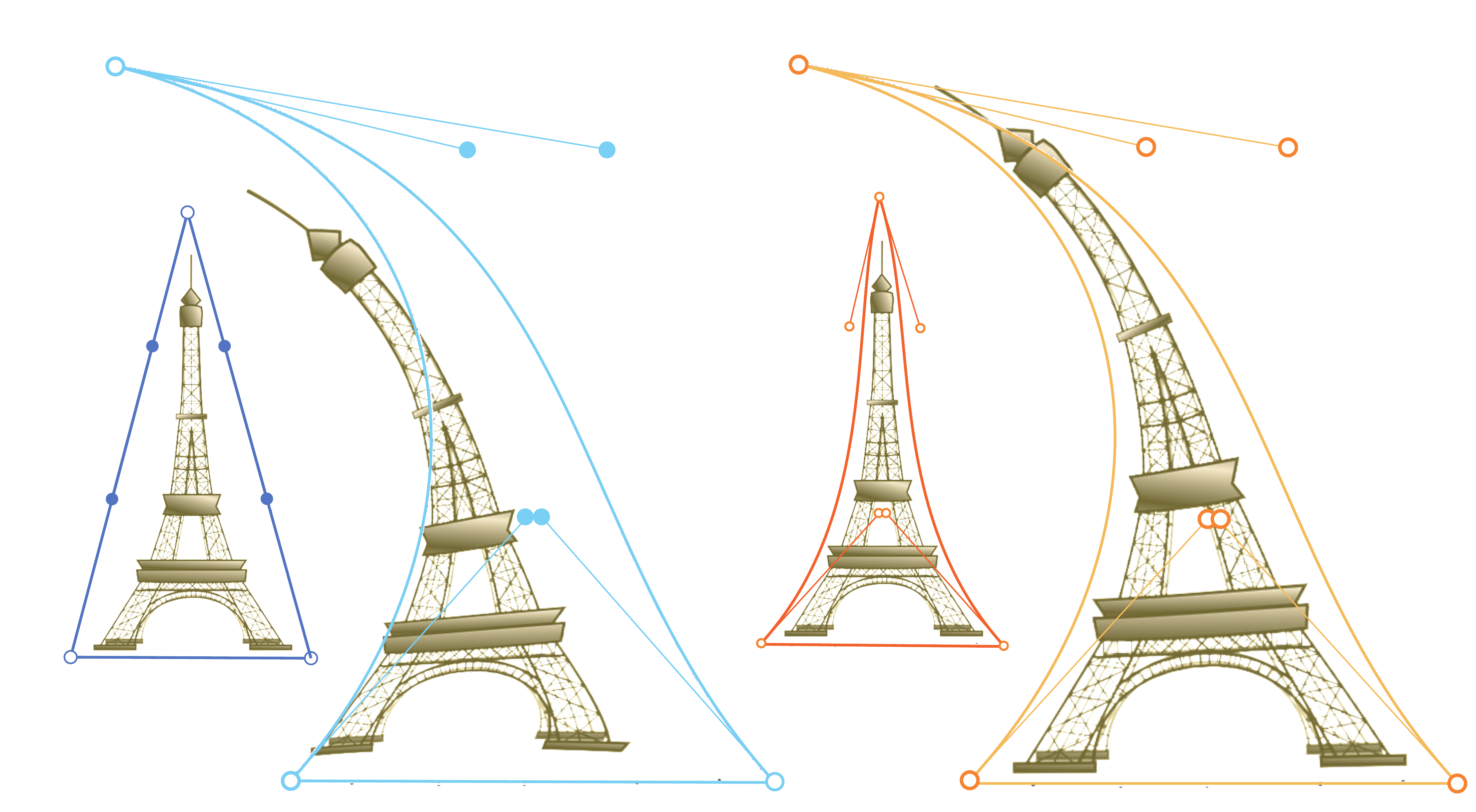} %
    {
    \put(6.5,6.5){\small Rest pose}
 \put(53,6.5){\small Rest pose}
    \put(12,-2){\small Polynomial Cauchy coordinates}
 \put(78,-2){\small Ours}
    }
  \end{overpic}
  \vspace{-1mm}
  \caption{
  Comparison with polynomial Cauchy coordinates, where the input linear (left) and high-order (right) cages have the same number of control points and are deformed to cubic cages. 
  }
  \label{fig:cmp-pgc}
\end{figure}

\paragraph{Comparisons to~\cite{Li2013}}
Cubic mean-value coordinates (Cubic MVCs)~\cite{Li2013} allow for deforming the straight cage edges into cubic curves.
Since linear cages are a special case of high-order cages, our coordinates can also be used for straight cages to achieve conformal deformation.
We obtain different conformal deformation results by assigning different non-linear parameterizations to each edge of the linear input cage (Fig. \ref{fig:cmp-mvc}).
However, the result of Cubic MVC has high distortion.

\paragraph{Comparisons to~\cite{Lin2024PolynomialCC}}
Given the same input linear and deformed high-order cages, polynomial Cauchy coordinates~\cite{Lin2024PolynomialCC} and polynomial Green coordinates~\cite{MichelThiery2023} generate the same deformation. Thus, we only compare polynomial Cauchy coordinates~\cite{Lin2024PolynomialCC} for deformation using straight input cages.
The number of control points is the same for polynomial Cauchy coordinates and ours for fair comparison.
In their deformation, since the control points of a deformed curve are on a straight line in the rest pose, the input linear cage may lack sufficient degrees of freedom to conform to input shapes closely.
%
When the linear input cage does not conform to the boundaries of the complex shape, the non-smoothness near the deformed boundaries exists (Fig.~\ref{fig:cmp-smoothness}), and the deformed shape is closer to the deformed cage using the higher-order cage via our coordinates than theirs (Fig.~\ref{fig:cmp-pgc}).
Thus, artists can achieve smoother and intent-matching deformations using high-order cage manipulation.
Finally, in Fig.~\ref{fig:comparep2p_poly}, we compare polynomial Cauchy coordinates using P2P deformation while the input cages have the same degrees of freedom.
In these cases, their linear cage does not closely conform to input shapes (e.g., no separation between the Elephant's hind legs in Fig.~\ref{fig:comparep2p_poly}), resulting in unnatural P2P deformations. 
High-order cages avoid this issue as they can tightly enclose the shapes.

\begin{figure}[t]
  \centering
  \begin{overpic}[width=0.965\linewidth]{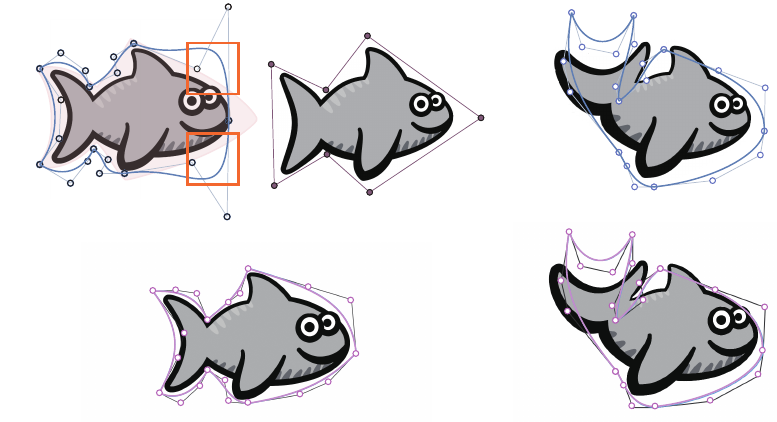} %
    {
     \put(2,24){\small High-order input cage}
    \put(36,24){\small Intermediate cage}
    \put(18,-2){\small High-order input cage}
    \put(62,25.5){\small Polynomial Cauchy coordinates}
    \put(84,-2){\small Ours}
    }
  \end{overpic}
  \vspace{-1mm}
  \caption{
  Comparison with polynomial Cauchy coordinates on high-order input cages. Their method requires two cages as input, needs a posteriori adjustments, and lacks a closed form (upper), whereas ours only needs one, enables direct deformation, and is closed-form (bottom). 
  }
  \vspace{-2mm}
  \label{fig:cmp-poly-Cauchy}
\end{figure}

Polynomial Cauchy coordinates can be used for high-order input cages. 
Specifically, they apply polynomial Cauchy coordinates on intermediate cages to generate high-order input cages for enclosing input shapes (Fig.~\ref{fig:cmp-poly-Cauchy}).
Then, they use numerical methods to compute the inverse of the mapping from the intermediate to the input cages to obtain the shape deformation.
Our coordinates have the following advantages.
First, their high-order input cage generation method cannot produce arbitrarily valid high-order input cages since the polynomial Cauchy coordinates are non-interpolating.
For input shapes with narrow regions (e.g., the area between the Octopus's claws or between the Elephant's hind legs in Fig.~\ref{fig:comparep2p_poly}), it is difficult to adjust their high-order input cages to avoid self-intersections, losing bijectivity to cause their algorithm failure.
Moreover, making their high-order input cages precisely conform to the shape is almost impossible.
%
Second, since their coordinates are not closed-form, their computation has larger errors and takes longer (see Table 1 in~\cite{Lin2024PolynomialCC}) and cannot be used for P2P deformation.

\section{Conclusion and Discussion}\label{sec:conclusion}
We propose polynomial 2D Cauchy coordinates for closed high-order cages containing polynomial curves of any order.
Our coordinates are obtained by extending the classical 2D Cauchy coordinates, enabling the transformation between polynomial curves of any order.
When applying our coordinate to 2D cage-based deformation, users can quickly generate the desired conformal deformation either by manipulating the cage's control points or by employing a point-to-point deformation approach. 
%
%
%
We demonstrate the effectiveness and practicability of our coordinates by extensively testing them on various 2D deformations. 



\paragraph{3D cages}
Our coordinates only work for 2D cages.
In computer graphics, 3D models are more widely used.
Since automatically and robustly generating 3D cages already exists~\cite{Guo2024RobustCage}, it is exciting to generalize our coordinates to 3D.


\paragraph{Conformal deformations}
Since conformality and interpolation conflict, our shape deformations do not always closely follow the deformed cage, as shown in the previous figures.
This may increase the difficulty for users in editing.
One possible way to reduce those conflicts is to allow users to specify regions where conformal properties can be relaxed.
Although there are solutions for the case where the input and output are all straight-edge cages~\cite{Lipman2008}, the method for high-order cages needs more exploration.




\section*{Acknowledgments}\label{sec:acknowledgments}
We would like to thank the anonymous reviewers for their constructive suggestions and comments.
This work is partially supported by the National Natural Science Foundation of China (62272429, 62025207). 

\begin{figure*}[t]
  \centering
  \begin{overpic}[width=0.99\linewidth]{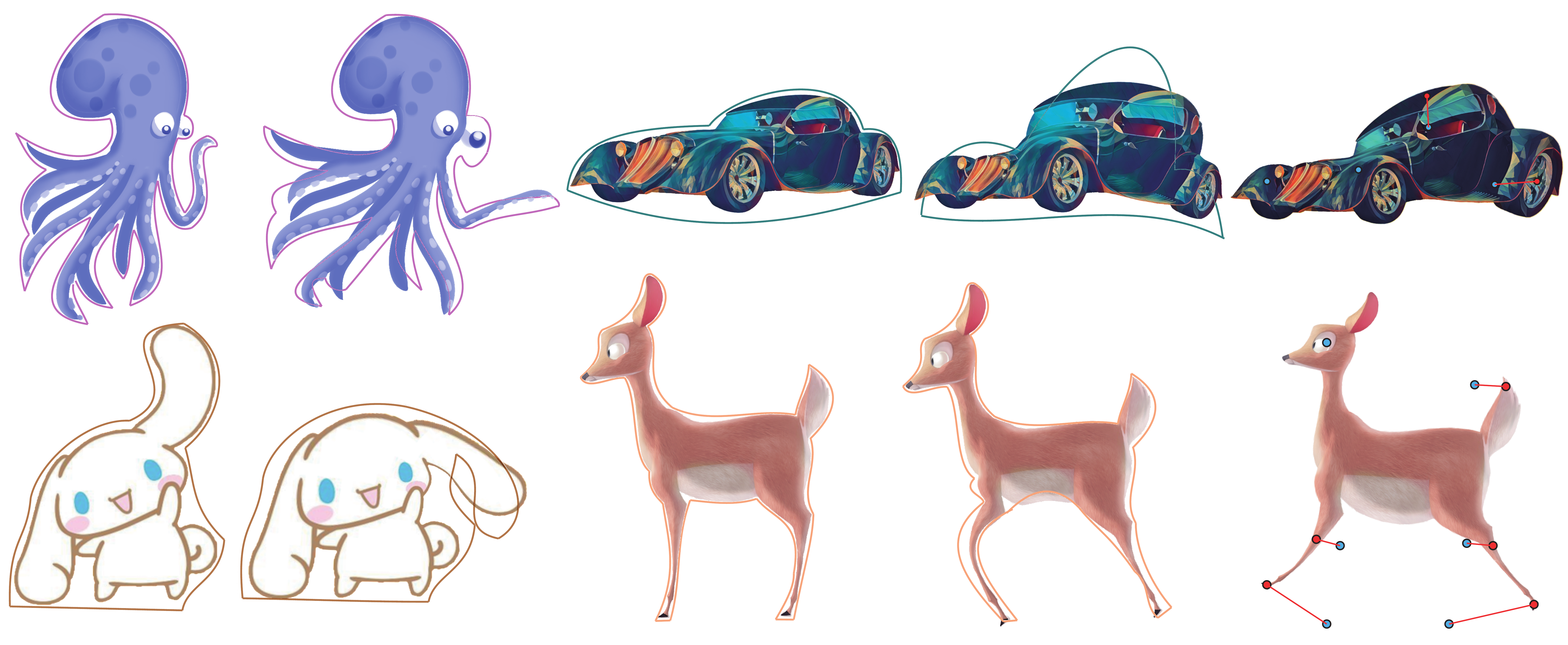} %
    {
    }
  \end{overpic}
  \vspace{-2mm}
  \caption{
Cage-based and P2P deformations using our coordinates and their derivatives for high-order input cages.
  }
  \label{fig:more-example}
\end{figure*}

\begin{figure*}[t]
  \centering
  \begin{overpic}[width=0.99\linewidth]{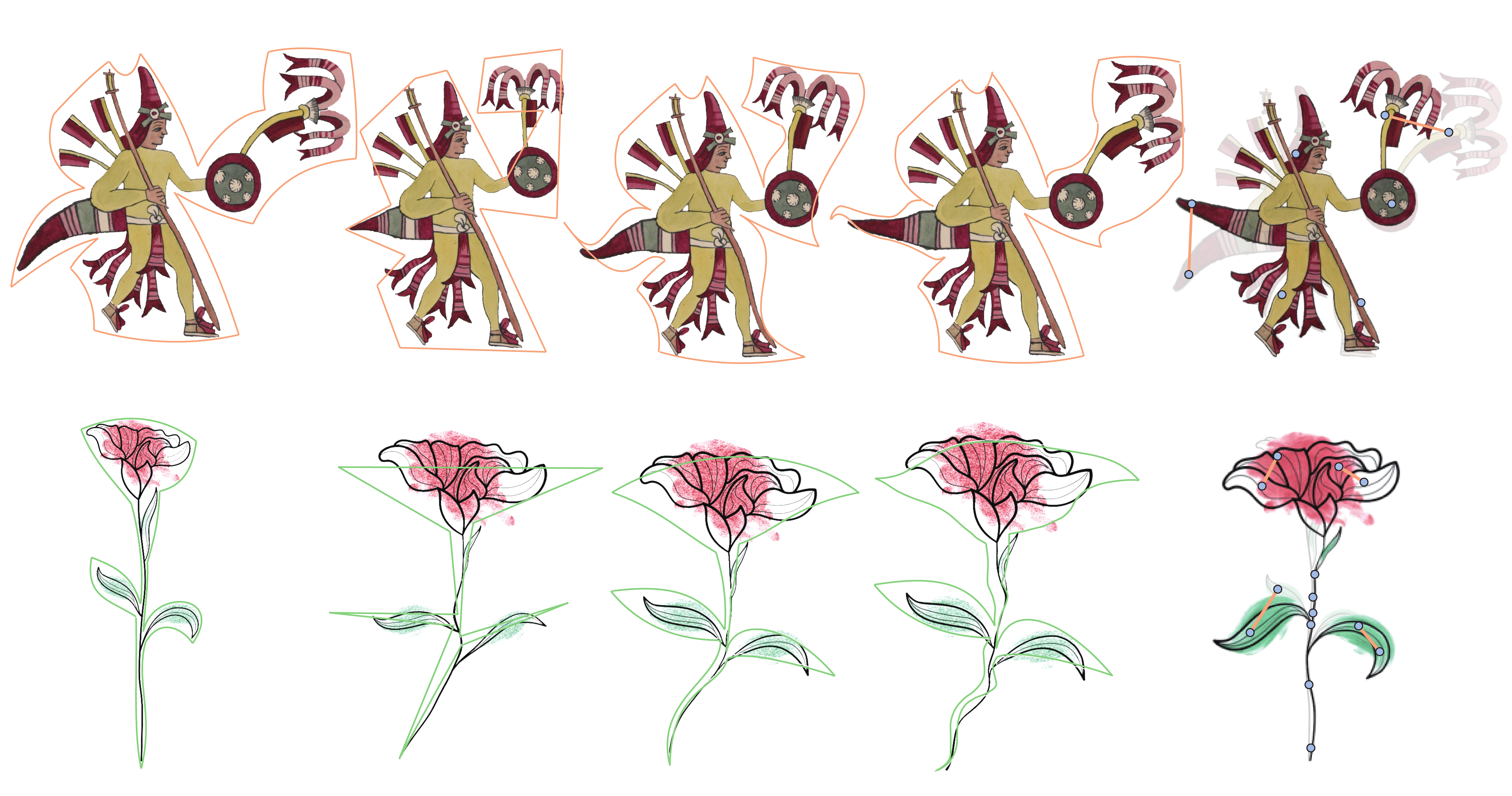} %
    {
    \put(9,-0.5){\small   Pose}
 \put(27,-0.5){\small Degree 1}
 \put(45,-0.5){\small Degree 3}
 \put(63,-0.5){\small Degree 7}
 \put(83,-0.5){\small Ours P2P}
    }
  \end{overpic}
  \vspace{-2mm}
  \caption{
  The high-order cages are deformed into high-order cages of different degrees.
  The degree of the input cage is two for the upper and three for the bottom.
  }
  \label{fig:diff-order}
\end{figure*}

\begin{figure*}[t]
  \centering
  \begin{overpic}[width=0.99\linewidth]{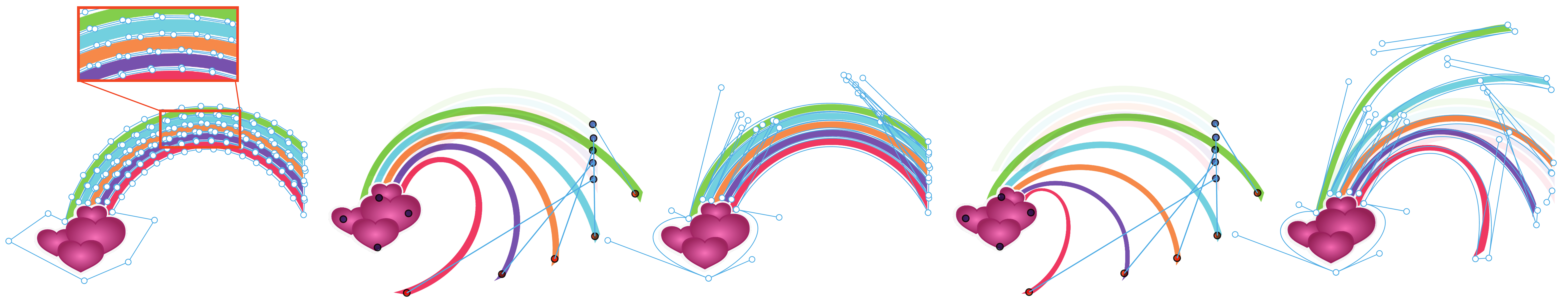} %
    {
    \put(1,-1.5){\small Linear input cage (151 points)}
 \put(43,-1.5){\small Cubic input cage (41 points)}
    \put(22,-1.5){\small P2P via Cauchy coordinates}
 \put(64,-1.5){\small P2P via our coordinates}
 \put(82,-1.5){\small Our cage-based deformation}
    }
  \end{overpic}
  \caption{
    Deformations using linear and high-order input cages. A linear cage requires 151 segments to enclose the given shape, particularly needing a very high number of segments in highly curved regions.
    The high number of segments makes editing prohibitively difficult, so we omit its cage-based deformation result here.
    A cubic cage, however, only needs 13 cubic curves and 5 straight lines to enclose the input shape. This significantly reduced number of parameters allows for direct editing of the cage to achieve deformations using our coordinates (rightmost).}
  \label{fig:cmp-linearcages}
\end{figure*}

\begin{figure*}[t]
  \centering
  \begin{overpic}[width=0.99\linewidth]{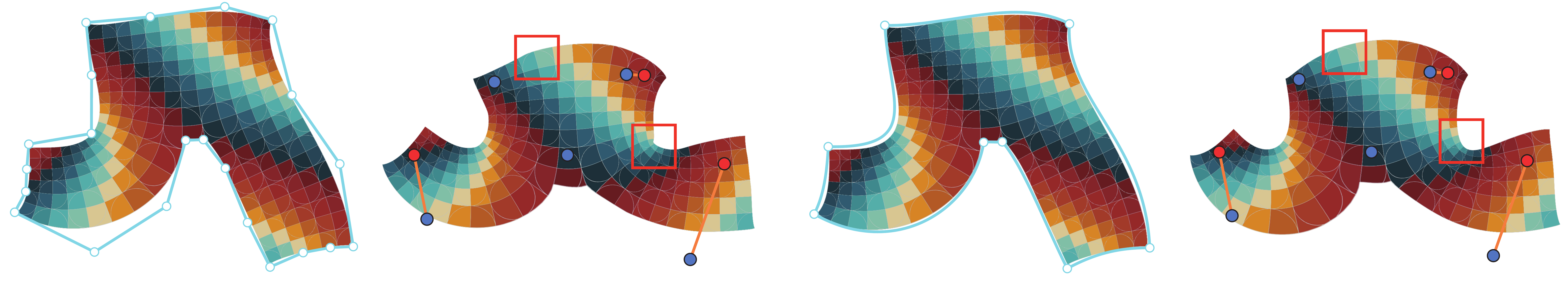} %
    {
    \put(6,0){\small Linear input cage}
 \put(56,0){\small High-order input cage}
    \put(29,0){\small Cauchy coordinates}
 \put(86,0){\small Ours}
    }
  \end{overpic}
  \vspace{-2mm}
  \caption{
  Comparison with Cauchy coordinates using the same number of control points for P2P deformations. The red boxes highlight the smoothness comparisons.
  }
  \label{fig:comparep2p}
\end{figure*}

\begin{figure*}[t]
  \centering
  \begin{overpic}[width=0.99\linewidth]{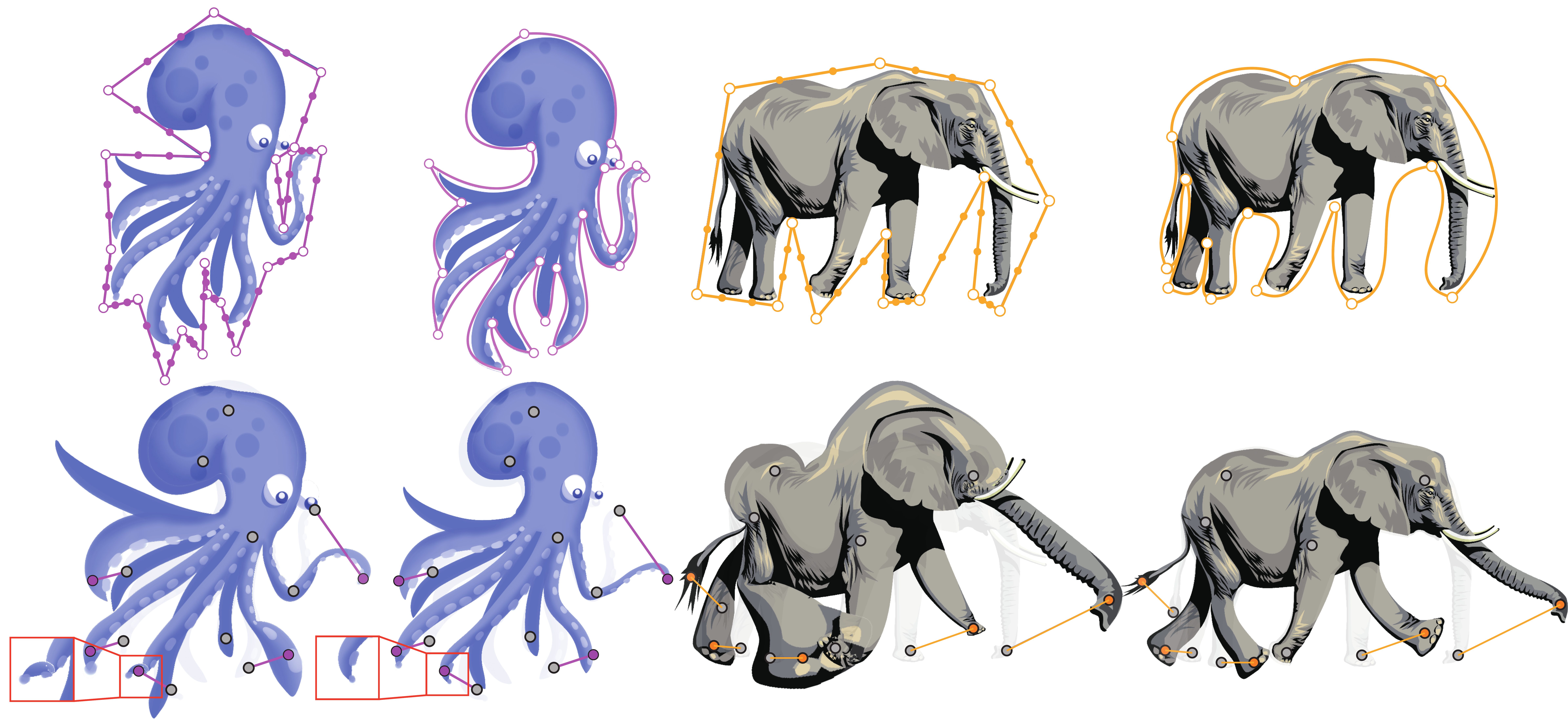} %
    {
    \put(3,-2){\small Polynomial Cauchy coordinates } 
 \put(32,-2){\small Ours}
  \put(45,-2){\small Polynomial Cauchy coordinates} 
 \put(83,-2){\small Ours}
    }
  \end{overpic}
  \vspace{1mm}
  \caption{
  Comparisons with polynomial Cauchy coordinates for P2P deformation, where the input cages (upper row) have the same number of control points. 
  In the linear input cages for polynomial Cauchy coordinates, non-endpoint control points on edges are displayed as solid dots. 
  }
  \label{fig:comparep2p_poly}
\end{figure*}

 
 
 
\ifCLASSOPTIONcaptionsoff
  \newpage
\fi

\bibliographystyle{IEEEtran}
\bibliography{src/reference}
%

\begin{IEEEbiography}[{\includegraphics[width=1in,height=1.25in,clip,keepaspectratio]{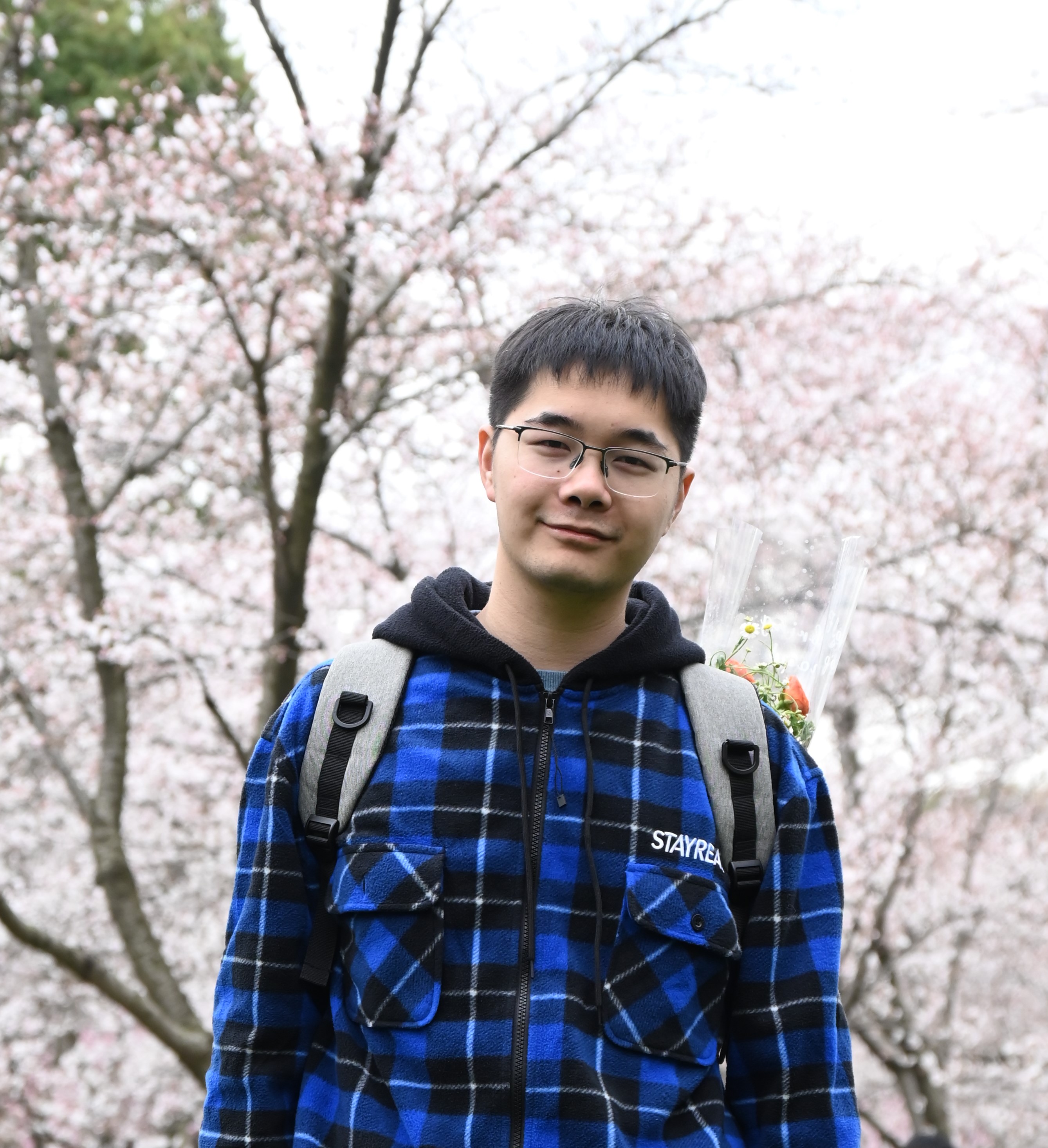}}]{Shibo Liu}
received a BSc degree in 2021 from the University of Science and Technology of China. He is currently a PhD candidate at the School of Mathematical Sciences, University of Science and Technology of China. His research interests include high-order geometric processing and 3D modeling. His research work can be found at his research website: \url{https://liu43.github.io/}.
\end{IEEEbiography}

\begin{IEEEbiography}[{\includegraphics[width=1in,height=1.25in,clip,keepaspectratio]{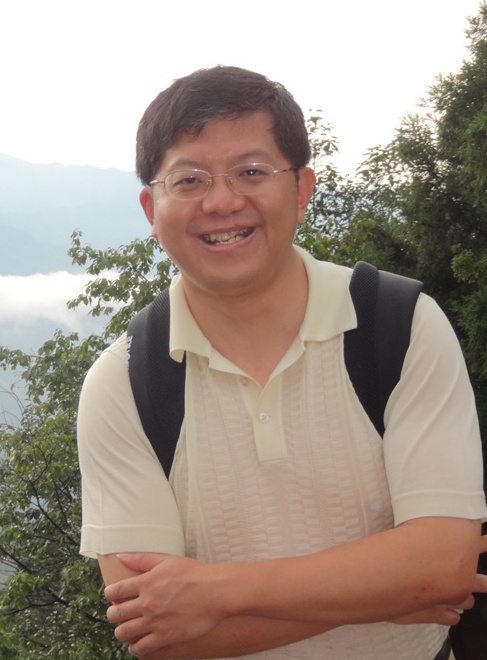}}]{Ligang Liu}
is a Professor at the School of Mathematical Sciences, University of Science and Technology of China. His research interests include computer graphics and CAD/CAE. His work on light-weight designing for fabrication at Siggraph 2013 was awarded as the first Test-of-Time Award at Siggraph 2023. \url{http://staff.ustc.edu.cn/~lgliu}.
\end{IEEEbiography}

\begin{IEEEbiography}[{\includegraphics[width=1in,height=1.25in,clip,keepaspectratio]{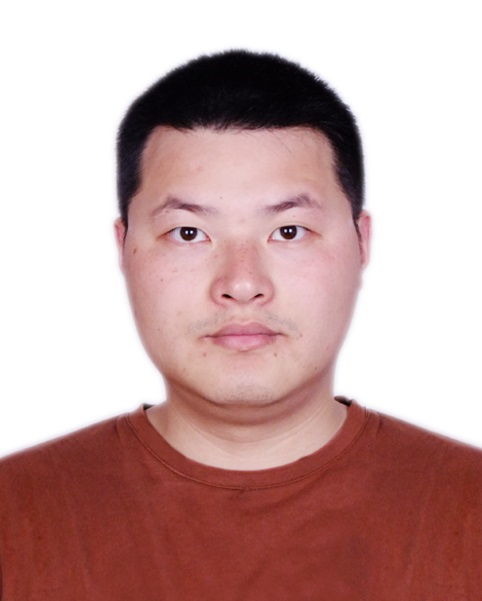}}]{Xiao-Ming Fu}
received a BSc degree in 2011 and a PhD degree in 2016 from the University of Science and Technology of China. He is an associate professor at the School of Mathematical Sciences, University of Science and Technology of China. His research interests include geometric processing and computer-aided geometric design. His research work can be found at his research website: \url{https://ustc-gcl-f.github.io/}.
\end{IEEEbiography}

\end{document}